# Temporally and Longitudinally Tailored Dynamic Space-Time Wave Packets

Xinzhou Su[1], Kaiheng Zou[1], Huibin Zhou[1], Hao Song[1], Yingning Wang[1], Ruoyu Zeng[1], Zile Jiang[1], Yuxiang Duan[1], Maxim Karpov[2], Tobias J. Kippenberg[2], Moshe Tur[3], Demetrios N. Christodoulides[1], and Alan E. Willner[1]

[1]Department of Electrical and Computer Engineering, Univ. of Southern California, Los Angeles, CA 90089, USA.

[2]École Polytechnique Fédérale de Lausanne (EPFL), Lausanne, Switzerland

[3]School of Electrical Engineering, Tel Aviv University, Ramat Aviv 69978, Israel

Corresponding emails: X.S. (xinzhous@usc.edu) or A.E.W. (willner@usc.edu)

**Abstract**

In general, space-time wave packets with correlations between transverse spatial fields and temporal frequency spectra can lead to unique spatiotemporal dynamics, thus enabling control of the instantaneous light properties. However, spatiotemporal dynamics generated in previous approaches manifest themselves at a given propagation distance yet not arbitrarily tailored longitudinally. Here, we propose and demonstrate a new versatile class of judiciously synthesized wave packets whose spatiotemporal evolution can be arbitrarily engineered to take place at various predesigned distances along the longitudinal propagation path. Spatiotemporal synthesis is achieved by introducing a 2-dimensional spectrum comprising both temporal and longitudinal wavenumbers associated with specific transverse Bessel-Gaussian fields. The resulting spectra are then employed to produce wave packets evolving in both time and axial distance – in full accord with the theoretical analysis. In this respect, various light degrees of freedom can be independently manipulated, such as intensity, polarization, and transverse spatial distribution (e.g., orbital angular momentum). Through a temporal-longitudinal frequency comb spectrum, we simulate the synthesis of the aforementioned wave packet properties, indicating a decrease in relative error compared to the desired phenomena as more spectral components are incorporated. Additionally, we experimentally demonstrate tailorable spatiotemporal fields carrying time- and longitudinal-varying orbital angular momentum, such that the local topological charge evolves every ~1 ps in the time domain and 10 cm axially. We believe that our space-time wave packets can significantly expand the exploration of spatiotemporal dynamics in the longitudinal dimension, and potentially enable novel applications in ultrafast microscopy, light-matter interactions, and nonlinear optics.

# Introduction

Sculpting the transverse degrees of freedom (DOFs) of different temporal frequency spectra has opened up new opportunities for generating a wide range of unprecedented and surprising spatiotemporal phenomena [1–7]. Along these lines, various families of space-time wave packets (STWPs), produced from such correlated space-time fields, have been experimentally and theoretically investigated [1,2,4,5]. In such wave packet configurations, different spatial field profiles are assigned to different temporal frequencies – an aspect that allows the generation of specific spatiotemporal field structures as a result of interference [1,2,4,5].

The implementation of these concepts has resulted in the emergence of diverse spatiotemporal processes, encompassing new dynamic behaviors and the ability to manipulate group velocities [8–30]. Advances in structured light and ultrafast pulse shaping have also made it possible to experimentally control the temporal dynamics of transverse light states [21,22,31], as illustrated in Fig. 1(a). However, that as of now, existing approaches do not naturally lend themselves in precisely controlling and engineering the dynamic evolution of STWPs along the axial propagation direction.

On the other hand, in terms of static longitudinal control, there is ability to synthesize monochromatic optical beams with arbitrary propagation-varying characteristics by utilizing structured light fields [32]. Specifically, multiple light DOFs (e.g., intensity, polarization, and orbital angular momentum (OAM)) have been longitudinally manipulated in a non-dynamic way [33–39] (Fig. 1(b)). In this case, the structured time-invariant beam is appropriately engineered so as to exhibit a tailored longitudinal wavenumber spectrum that is correlated with the transverse fields. In view of these developments, it might be of great interest to devise new methodologies via which the dynamic evolution of prudently synthesized spatiotemporal wave packets can be precisely pre-engineered along the axial path.

In this paper, we synthesize a new class of STWPs with enriched spatiotemporal properties that can now be controllably varied in both time and longitudinal distance. This is accomplished by appropriately correlating the 2-dimensional (2-D) spectra associated with the temporal and longitudinal wavenumbers that are assigned to different transverse Bessel-Gaussian (BG) fields (Fig. 1(c)). By employing 2-D Fourier transform schemes, various light wave packet attributes can be independently controlled, such as intensity, polarization, and transverse spatial distribution (e.g., OAM). Simulation results of these behaviors show that utilizing more spectral components can reduce the relative error when compared to the desired spatiotemporal phenomena. Additionally, we experimentally demonstrate a STWP whose local OAM evolves along the time and distance coordinates. In this case, the on-axis topological charge is transformed every ~1 ps in time and/or 10 cm in the axial distance with an arbitrarily designed mode assignment. This versatile category of wave packets can offer new opportunities in unveiling and studying spatiotemporal phenomena, especially where the spatiotemporal light-matter interactions at particular distance is of crucial importance.

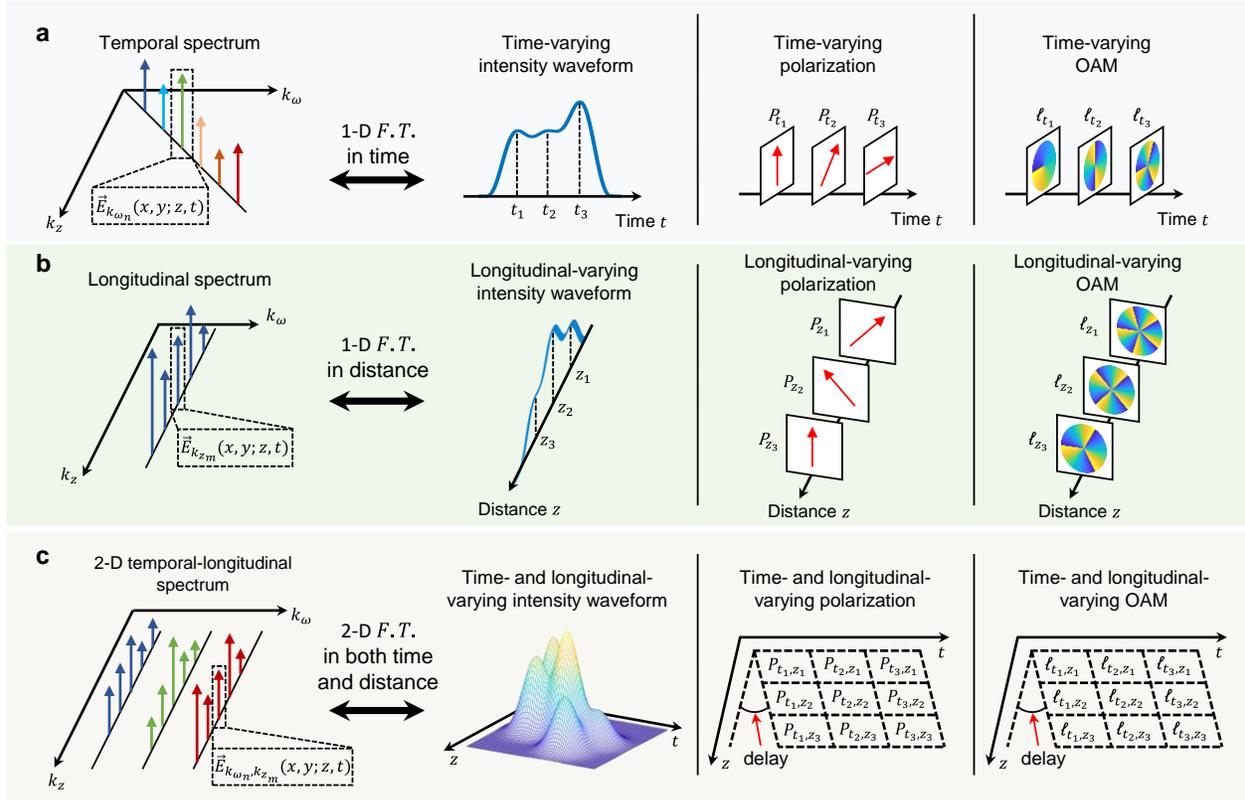

**Fig. 1. Concept of structured light fields with designed temporal and longitudinal evolutions.** (a) Coupled transverse spatial fields ($\vec{E}_{k_\omega}$) and temporal frequency spectra ($k_\omega$) can give rise to dynamic spatiotemporal phenomena. The temporal dynamics of light DOFs can be independently controlled (e.g., intensity, polarization, and OAM) through Fourier transforms (*F.T.*). With each temporal wavenumber correlated with a single longitudinal wavenumber, the spatiotemporal property is typically not tailored longitudinally. (b) By tailoring the transverse spatial fields ($\vec{E}_{k_z}$) of longitudinal wavenumbers ($k_z$) at the same temporal frequency, a light beam with longitudinally structured properties can be produced. (c) Through manipulating a 2-D spectrum consisting of both temporal and longitudinal wavenumbers associated with transverse spatial fields ($\vec{E}_{k_\omega,k_z}$), versatile time- and longitudinal-varying spatiotemporal phenomena can be generated. The wave packets manifest as parallelogram shapes in time-distance coordinates due to the propagation delay.

## Results

**Mechanism of temporal and longitudinal evolution** To intuitively explain how the 2-D temporal and longitudinal spectral space can lead to spatiotemporal phenomena in both time and distance, we illustrate an example of interference between several mutually coherent harmonic waves. The wave at a specific time instant $t$ and distance $z$ is described as [40]

$$E(z,t) = e^{i(k_z z - k_\omega c t)} \tag{1}$$

where $c$ is the speed of light in the vacuum; $k_\omega$ represents the temporal wavenumber and is directly linked to the angular temporal frequency $\omega$, such that $\omega = k_\omega c$; and $k_z$ is the longitudinal wavenumber.

Figure 2(a) shows the beating of the two temporal frequencies in the time and distance domains. At a given distance, different constructive or destructive interference between the two temporal frequencies as a function of time results in a periodic temporal amplitude envelope. The period $T$ corresponds to the temporal frequency difference $\Delta\omega$, and $T = 2\pi/\Delta\omega$. For a wave packet comprising multiple temporal frequencies, the relationship between $\omega$ and $k_z$, known as the dispersion relation, determines the group velocity expressed as $v_g = \partial\omega/\partial k_z$. This dispersion relation can be tailored to achieve arbitrary control over the group velocity [2,10,11]. When $k_z$ is linearly correlated with $k_\omega$ by a factor $\alpha$ (i.e., $\partial k_\omega/\partial k_z = \alpha$), the group velocity of the wave packet is derived as $v_g = \alpha \cdot c$. As a result, a time delay in the temporal envelope, indicating propagation, can be observed along the axial propagation direction.

Similarly, a longitudinally varying interference pattern can be observed when coherently combining two longitudinal wavenumbers at the same temporal frequency, as illustrated in Fig. 2(b). In this scenario, the longitudinal amplitude envelope remains stationary over time due to the shared single temporal carrier frequency. Analogously, the longitudinal period $L$ depends on the difference in wavenumbers $\Delta k_z$ between the two longitudinal spectral components, and thus, $L = 2\pi/\Delta k_z$.

Combining the manipulation in both spectral spaces, waveform control can be extended across both the time and distance domains. As depicted in Fig. 2(c), two temporal wavenumbers, each associated with two distinct longitudinal wavenumbers, are combined. The phase delay variation among different spectral components exhibits dependencies in both dimensions. This yields a periodically recurring parallelogram-shaped interference pattern within the 2-D time-distance coordinates. Temporal intensity waveform evolves along the axial direction. Alternatively, when observed from the longitudinal direction, the interference pattern also undergoes evolution over time. The slope of the parallelogram indicates the group velocity of the temporally and longitudinally structured wave packet.

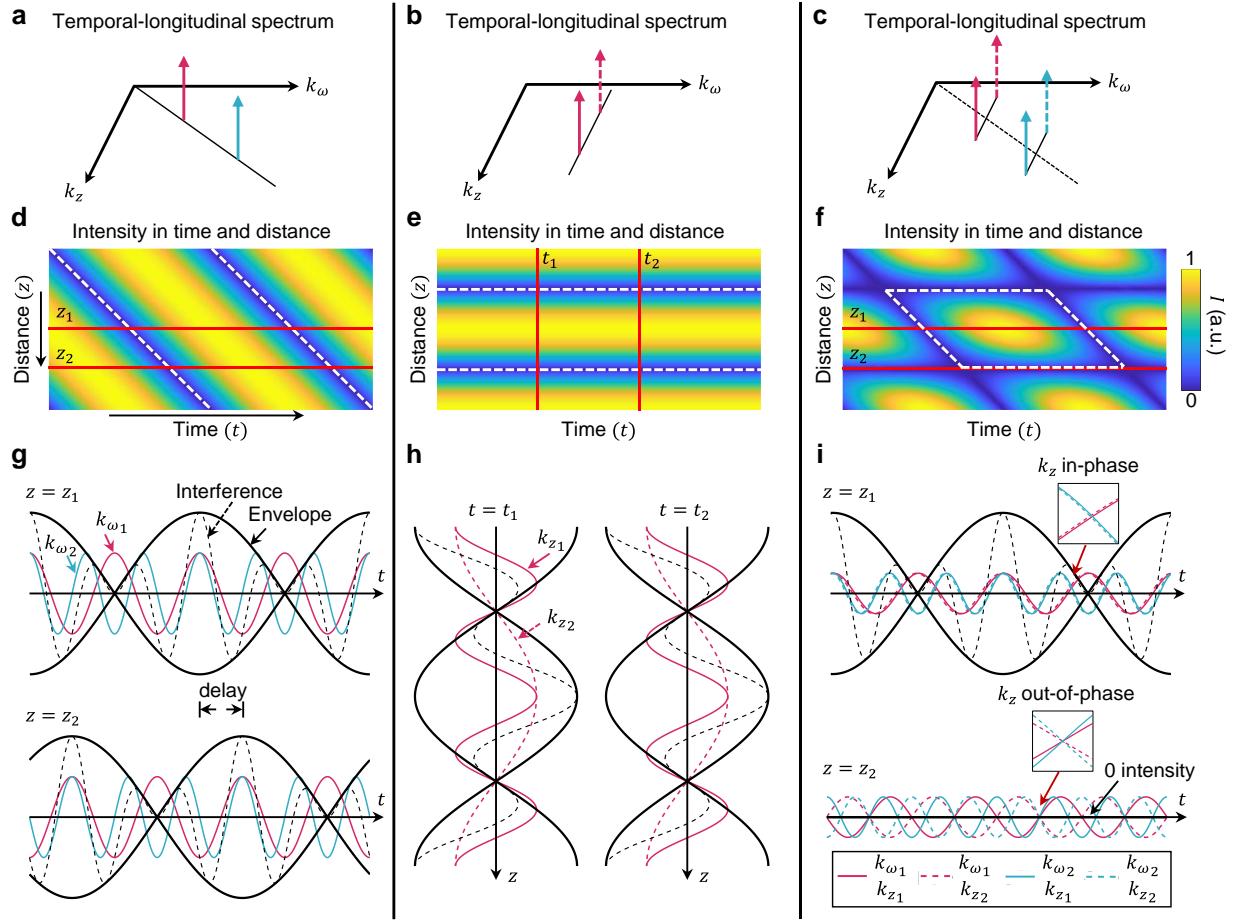

**Fig. 2. Beating between coherent harmonic waves.** (a) Superposition of two temporal frequencies, each associated with a longitudinal wavenumber. $k_z$ is linearly correlated with $k_\omega$, and the gradient determines the group velocity. (b) Superposition of two longitudinal wavenumbers at the same temporal frequency. (c) Superposition of two temporal frequencies, each carrying two distinct longitudinal wavenumbers. Four spectral components are evenly distributed across the spectrum. (d-f) The intensity of the superposed field with respect to time and distance. White dashed lines indicate the interference period. In (d) and (f), the intensity pattern is longitudinally displaced by propagation delay. (g-i) Waveform interference at different distance/time cross-sections in (d-f). In the waveform, the red and blue lines represent the real components of harmonic waves. The black dashed line shows the coherent sum, and the black solid line depicts the amplitude envelope.

**Designing temporally and longitudinally tailored STWPs** The fundamental harmonic wave beating that we present underlies the basic mechanism for understanding temporal and longitudinal evolution, disregarding transverse spatial distributions. In optical systems, an approximation of such interference can be realized within the paraxial regime by employing beams with tunable longitudinal wavenumbers. One example is using a BG modal basis. BG modes enable direct control of both $k_\omega$ and $k_z$ of a beam, and the electric field $\vec{BG}_{k_\omega,k_z,\ell}$ can be expressed as follows [41]:

$$\vec{BG}_{k_\omega,k_z,\ell}(r,\theta;z,t) = G(r) \cdot J_\ell(k_r r) e^{i\ell\theta} e^{i(k_z z - k_\omega c t)} \hat{\epsilon} \qquad (2)$$

where $r$, $\theta$, and $z$ are the radius, angle, and distance in a cylindrical coordinate system, respectively; $\ell$ is the OAM order of the Bessel mode; $k_r$ represents the transverse wavenumber, satisfying $k_r^2 + k_z^2 = k_\omega^2$; $\hat{\epsilon}$ denotes the polarization state; and $G(r)$ is a Gaussian apodization which determines the Rayleigh range of the beam, indicating the low divergence region for the BG beams.

Meanwhile, the harmonic wave beating shown in the previous case presents a basic oscillating interference pattern. To achieve versatile spatiotemporal phenomena controlled in both time and distance, it becomes necessary to design a 2-D temporal-longitudinal spectrum employing a greater range of spectral components. Here, we investigate a design method using the Fourier relation between the temporal and longitudinal waveforms and their corresponding spectrum. We define the waveform as the on-axis complex amplitude and phase in terms of time or distance. In one-dimensional (1-D) scenarios, arbitrary temporal/longitudinal waveforms are demonstrated by temporal/longitudinal spectral shaping through the 1-D Fourier transform [42,43]. Extending the concept to 2-D waveform construction, the required spectrum can be obtained by performing a 2-D Fourier transform. For the sake of simplicity, we apply a Galilean transform to the basis and define the time as $\tau = t - z/v_g$, thus eliminating the impact of propagation during the design process. Therefore, an arbitrary 2-D $z - \tau$ waveform can be synthesized through superposition:

$$s(z,\tau) = E(r=0, \theta=0; z, \tau) = \sum_n \sum_m C_{n,m} \cdot BG_{k_{\omega n,m}, k_{z n,m}, \ell=0}(r=0, \theta=0; z, \tau) \qquad (3)$$

where $s(z,\tau)$ describes the desired complex amplitude and phase waveform as a function of time and distance; $C_{n,m}$ is the complex coefficient of the $n$-th temporal and $m$-th longitudinal spectral component, calculated using the Fourier relation (*"Methods"*). Each frequency component maintains the same polarization and carries a fundamental $\ell = 0$ mode.

Furthermore, various DOFs (e.g., polarization, OAM) existing in the transverse fields provide opportunities for generating exotic time- and longitudinal-varying light states. In pursuit of such control, the electric field of the desired wave packet can be decomposed into multiple sub-wave packets, each having a unique transverse state. For instance, polarization evolution can be structured to evolve temporally and longitudinally. Given that any polarization state can be expressed via its Jones vector, an arbitrary time- and longitudinal-varying polarization (TLV-Pol) $\vec{E}_{TLV-Pol}$ can be broken down into two mutually orthogonal linearly polarized wave packets, $E_x$ and $E_y$ [35,36]:

$$\vec{E}_{TLV-Pol}(r,\theta;\tau,z) = \begin{bmatrix} E_x(r,\theta;\tau,z) \\ E_y(r,\theta;\tau,z) \end{bmatrix} \quad (4)$$

Providing a polarization evolution pattern, the amplitude and phase waveforms of $E_x$ and $E_y$ are determined and can be realized by temporal-longitudinal waveform shaping. Similarly, STWPs carrying time- and longitudinal-varying OAM (TLV-OAM) can be synthesized by combining sub-wave packets of different OAM orders. The superposed field $\vec{E}_{TLV-OAM}$ is given as [22,38,39]:

$$\vec{E}_{TLV-OAM}(r,\theta;z,\tau) = \sum_j \vec{E}_{\ell_j}(r,\theta;z,\tau) \quad (5)$$

where $\vec{E}_{\ell_j}$ is the sub-wave packet carrying OAM $\ell_j$, labeled by $j$. The OAM state at a specific time instant and distance is determined by the intensity distribution among different sub-wave packets at that specific position. We note that temporal-longitudinal response of these light properties is independently designed. Consequently, different DOFs can be simultaneously customized, achieving versatile control of the transverse states in time and distance domains.

**Manipulating temporal-longitudinal waveform** Using the proposed approach, we simulate temporally and longitudinally tailored STWPs with control of multiple light DOFs. In this demonstration, we discuss discrete and equally spaced frequency comb spectra, while continuous spectra can also be employed. Aligned with the experimental schemes detailed in the following sections, the temporal frequency spacing $\Delta f$ is set to be 192 GHz (which corresponds to $T = 5.2\ ps$), and the longitudinal wavenumbers at the same temporal frequency have a spacing $\Delta k_z$ of $2\pi/60$ cm$^{-1}$ (resulting in $L = 60\ cm$). Particularly, we design the spatiotemporal phenomena within a pulse width of 3.47 ps and longitudinal distance of 30 cm. The propagation distance is partitioned into three ranges, each spanning 10 cm, and temporal dynamics are designed separately for each longitudinal range.

We first show the manipulation of the temporal-longitudinal waveform of the wave packet. Three different temporal intensity waveforms are devised, corresponding to three distance ranges. For $0 \leq z < 10\ cm$, a rectangular waveform covers the entire temporal range; for $10 \leq z < 20\ cm$, two short rectangular pulses are designed with 0.86 ps pulse width; and For $20 \leq z \leq 30\ cm$, the temporal waveform follows a triangular pulse shape. The simulated on-axis intensity in a full 2-D $z - \tau$ space using 15 (temporal)×15 (longitudinal) wavenumbers indicates the intensity evolution across both temporal and longitudinal domains (Fig. 3(a)). At three central distances, temporal evolution of transverse intensity profiles along $x$-axis at $y = 0$ are shown in Fig. 3(b) and *Extended Data Fig. 1*, varying numbers of wavenumbers for the synthesis process. With a limited number of spectral components, the truncated spectrum causes an imperfect waveform generation. To evaluate the quality of the generated waveform, we define a relative intensity error as

$$error = \int \left| \frac{I_{ideal}(\tau) - I_{simulation}(\tau)}{I_{ideal}(\tau)} \right| d\tau \tag{6}$$

With increased spectrum resources, the relative error of the waveform decreases from 0.150 to 0.028, 0.384 to 0.095, and 0.089 to 0.011 at three distances, respectively (Fig. 3(c)). Additionally, the relative error strongly depends on the particular shape of the desired waveform. For example, a narrower pulse ($10 \leq z < 20 \ cm$) requires more higher frequency components to achieve a low error.

**Tailor polarization states in time and distance** In Fig. 3(d), we simulate a wave packet in which the on-axis polarization undergoes different dynamical evolution along axial distances. This phenomenon is achieved by manipulating amplitude waveforms for $\hat{x}$- and $\hat{y}$-polarized sub-wave packets, along with the adjustment of the phase difference between them. The simulated on-axis polarization states as a function of time are illustrated on the Poincaré sphere for three central distances. In *Supplementary Movie 1*, we also present the polarization distribution evolution across the whole beam profile. For $0 \leq z < 10 \ cm$, the polarization state evolves from $\hat{x}$ polarization to $\hat{y}$ polarization along the equatorial plane; for $10 \leq z < 20 \ cm$, the state transitions between circular polarizations and follows a trajectory along a meridian; and for $20 \leq z \leq 30 \ cm$, the trajectory combines both meridian and equatorial paths. The simulated amplitude waveforms of $E_x$ and $E_y$, and the phase difference $\varphi(E_y) - \varphi(E_x)$ show good agreement with the design (Fig. 3(e) and *Extended Data Fig. 2*). The instantaneous polarization state is expected to be closer to the devised state when more spectra components are employed for the synthesis.

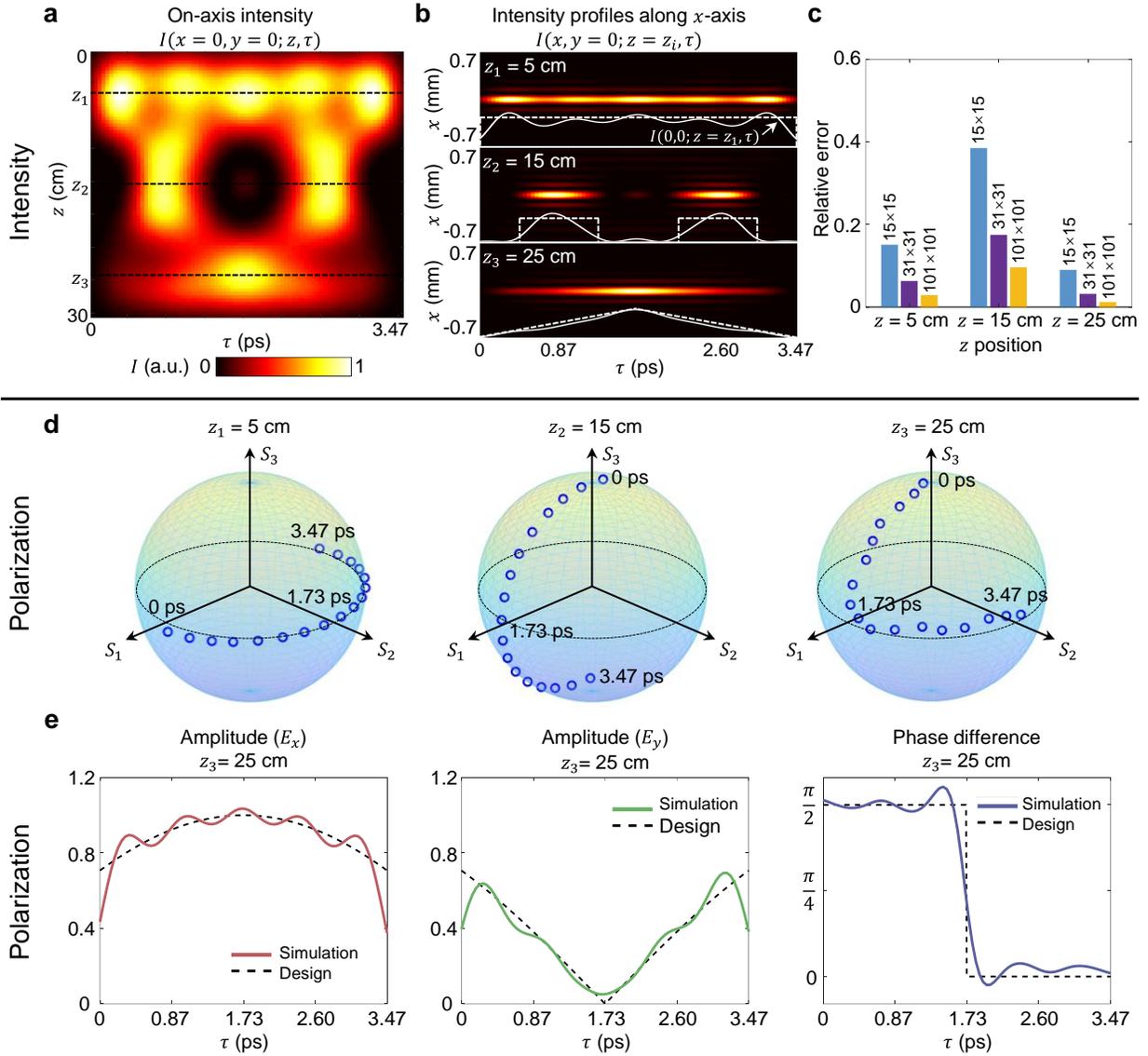

**Fig. 3. Simulated STWP structured in both time and axial distance.** (a-c) STWP with temporal-longitudinal waveform control. (a) On-axis intensity evolution $I(x=0, y=0; z, \tau)$ in time-distance coordinates. The wave packet is synthesized using 15 (temporal)×15 (longitudinal) spectral components. (b) Temporal evolution of intensity profiles along $x$-axis $I(x, y=0; z=z_i, \tau)$ at three central distances. The solid white line represents the on-axis intensity, and the dashed white line denotes the devised waveform. (c) Relative intensity error at different central distances with varying numbers of frequencies employed for wave packet synthesis. (d-e) A different STWP generated with TLV-Pol. (d) Poincaré sphere representation shows simulated instantaneous on-axis polarization states in terms of time at three central distances. 15×15 frequencies are utilized. (e) Ideal and simulated amplitude waveforms of $E_x$ and $E_y$ and phase difference $\varphi(E_y) - \varphi(E_x)$ at $z_3 = 25$ cm.

**Synthesizing STWPs carrying TLV-OAM** We also theoretically investigate and experimentally demonstrate STWPs carrying TLV-OAM. This type of pulse is constructed by superposing multiple sub-STWPs each carrying a distinct topological charge value. Our approach allows for the arbitrary allocation of OAM values in time and distance domains.

To experimentally synthesize TLV-OAM wave packets, we tailor the spatiotemporal light field from a frequency comb source, as detailed in *Supplementary Note 1*. A Kerr frequency comb is directed into a programmable liquid crystal on silicon (LCoS) filter, which selects multiple frequency lines for manipulation. Different frequency lines are separated and projected to distinct locations on a spatial light modulator (SLM). The longitudinal wavenumbers are imparted by modulating the beam with the corresponding transverse spatial fields. Thus, all longitudinal components are coherently combined and subsequently encoded to the temporal frequency all at once. Finally, different temporal frequency components are co-axially combined to construct the wave packet ("*Methods*"). To characterize the spatiotemporal field of the generated STWP, we develop a spatiotemporal off-axis holographic setup. Two translation delay stages are set on both arms of the generated STWP and the reference pulse. The translation of the STWP path ($\Delta z_{STWP}$) is used to control the propagation distance, and the translation of the reference path ($\Delta z_{ref}$) determines the time of the STWP. The complete electric field of the STWP $E_{STWP}(x, y; z, \tau)$ at a specific distance $z$ and time $\tau$ can be reconstructed through the off-axis holography as $E_{off-axis}(x, y; z = \Delta z_{STWP}, \tau = \Delta z_{ref}/c - \Delta z_{STWP}/v_g)$ ("*Methods*").

We simulate and experimentally generate STWPs with different OAM value assigned at 4 time instants and 3 distances synthesized by 6 (temporal) × 11 (longitudinal) spectral components. To make full use of the optical power on each temporal frequency line, we design the temporal waveform as Sinc functions, corresponding to a rectangular spectrum in the temporal frequency domain. In this particular demonstration, the topological charge of the STWP changes every 0.87 ps in time and every 10 cm in distance. Figures 4(a, c) illustrate the simulated and experimental intensity and phase profiles at the center of these time and distance slots. The OAM value displays an increment of +1 between consecutive slots. At the beam center, the electric fields with the designated topological charge constructively interfere. In contrast, the other OAM fields destructively interfere, and the power is distributed outside the central lobe, which is indicated by the spiral pattern in the intensity and phase profiles. In these wave packets, the spin and orbital angular momenta experience energy change between central and side lobes while the total momenta of the system is conserved [38,39]. Figures 4(b, d) show the temporal evolution of the intensity profiles along $x$-axis at $y = 0$ at three central distances. The full transverse spatial evolution at a fixed distance or time is available in *Supplementary Movie 2-3*.

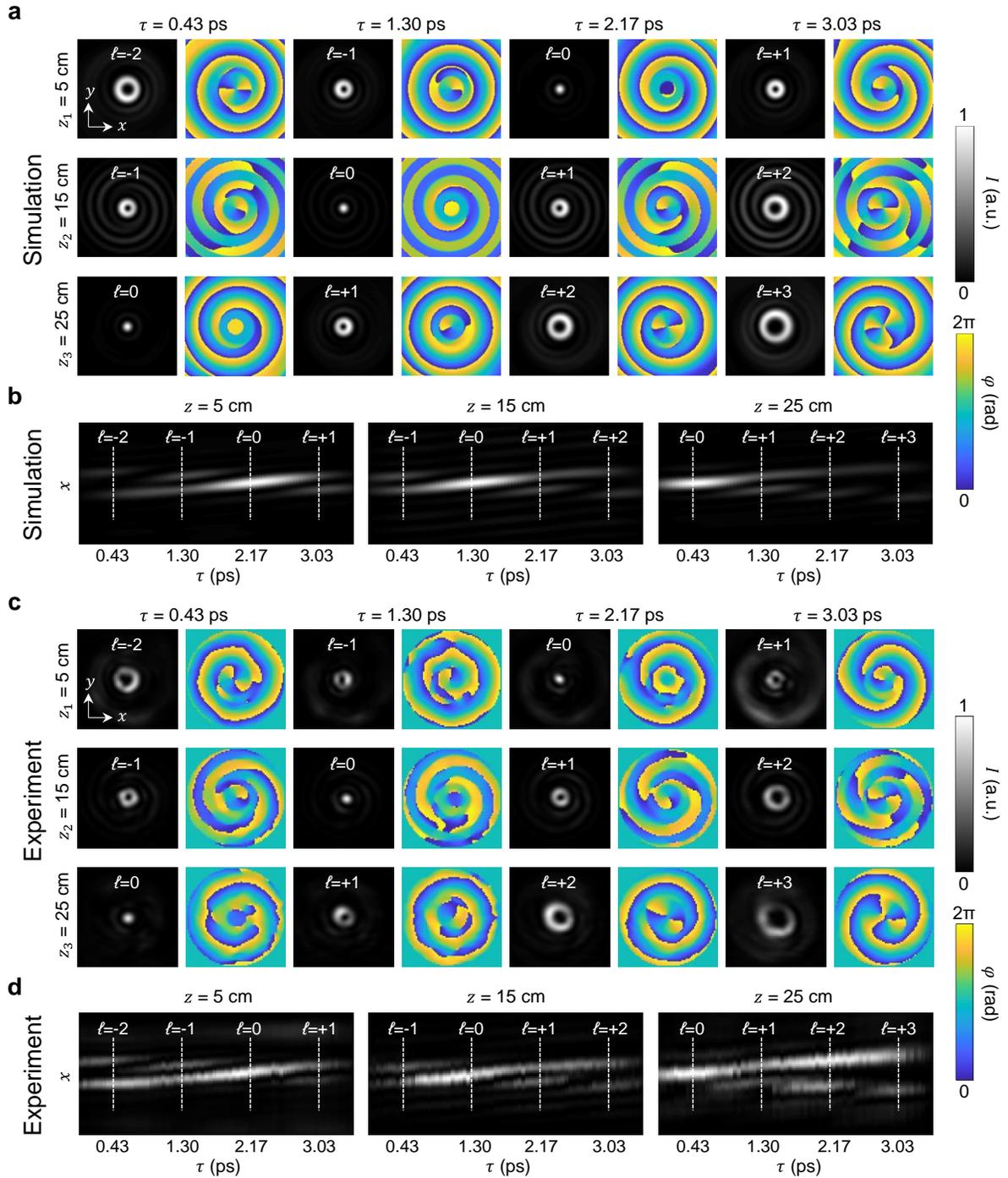

**Fig. 4. Profiles of a STWP carrying TLV-OAM.** (a) Simulated intensity and phase profiles at different time instants and distances. The OAM value is designed to increase by +1 with time/distance. The profiles have dimensions of 1.4 mm × 1.4 mm. (b) Temporal evolution of intensity profiles along $x$-axis $I(x, y = 0; z = z_i, \tau)$ at three distances. (c) Experimentally measured intensity and phase profiles of the same TLV-OAM wave packet. (d) Temporal evolution of experimental intensity profiles along $x$-axis.

To characterize the beam quality of the generated wave packet, we calculate the modal purity at the center of the spatial field. A limited-size aperture is used to cover the central lobe of the transverse field. With the same mode assignment, we compute the dynamic evolution of OAM purity at three central distances (Fig. 5(a)). The modal purity at the center of each time slot ranges from ~75% to ~90%. Such imperfection can result from impairments in transverse field generation, fluctuation in power and phase among temporal frequencies, etc. The modal power distribution during the transition stage clearly indicates the superposition of different OAM states. We also calculate the modal purities given a fixed time and provide simulated modal purities with the same configuration as that presented in *Extended Data Fig. 3*. For comparison, a STWP carrying time-varying OAM is also simulated in *Supplementary Note 1*. In contrast to TLV-OAM wave packets, the same temporal evolution behavior of OAM values replicates along the longitudinal distance.

Our spatiotemporal synthesis approach also provides the flexible tunability in the assignment of OAM values. By simply substituting the mode order allocated at different time instants or distances while maintaining identical complex coefficients for the spectrum, we can easily adjust the temporal and longitudinal OAM variations. As shown in Figs. 5(b-d), we experimentally demonstrate multiple TLV-OAM wave packets and their local topological charge change in diverse manners: OAM values that consistently decreases in time and axial distance; the OAM value change following different dynamics at three distances, including increments, decrements, and sweeping between modes; and exhibiting discontinuous mode shifts. The corresponding modal purities are calculated. In particular, transverse field profiles and simulated purities for the case in Fig. 5(c) are presented in *Extended Data Figs. 4 and 5*. It is worth noting that, in this demonstration, we employ discrete and equally spaced time and distance intervals to assign OAM values, while the position and width of such intervals can also be flexibly designed [20].

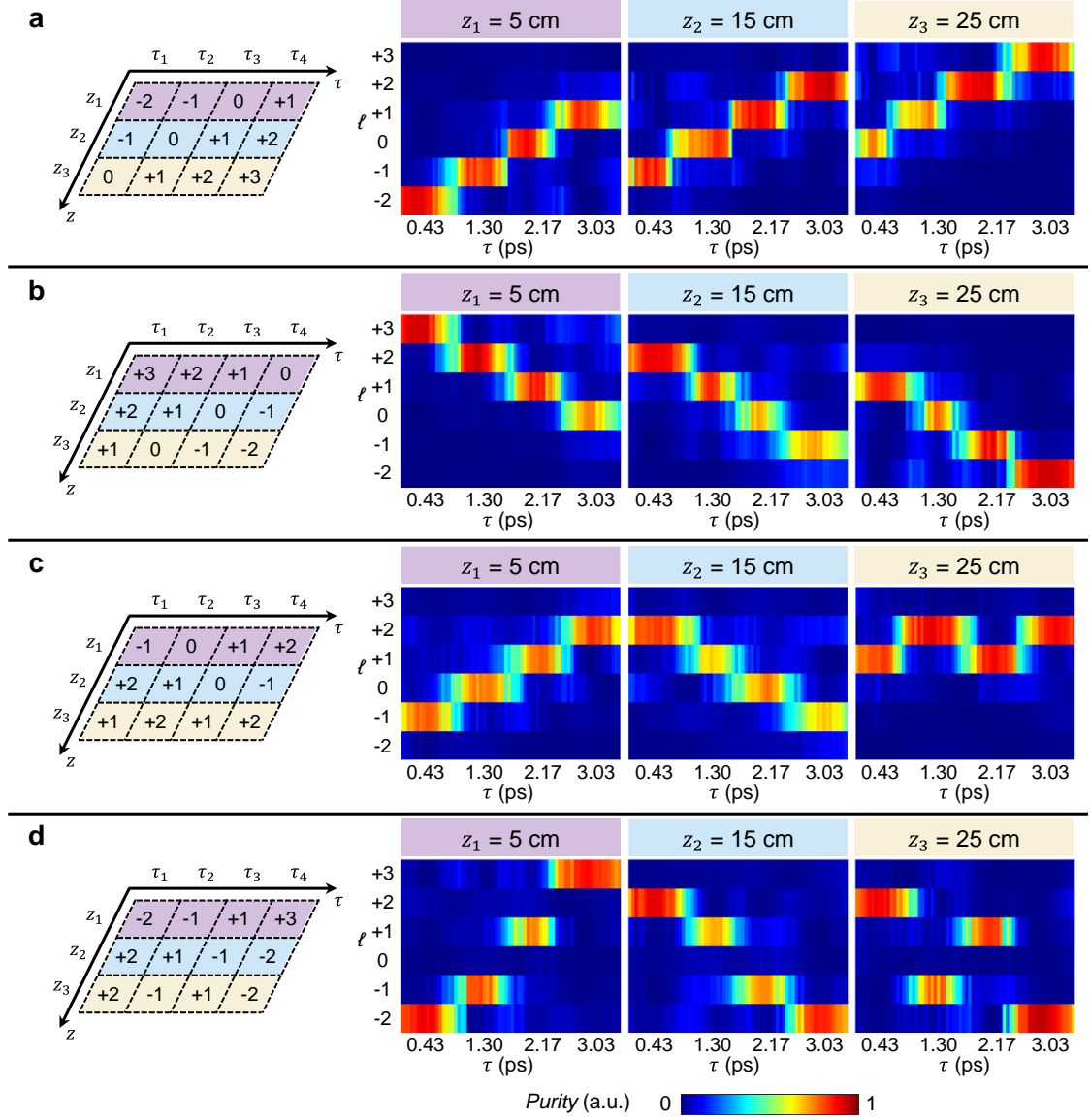

**Fig. 5. Color coded modal spectra for various TLV-OAM wave packets.** (a-d) Experimental OAM modal spectra in terms of time at three central distances. An aperture with a radius of 0.3 mm is applied to filter the central lobe of the transverse fields. Multiple realizations of the OAM evolving pattern are generated experimentally: (a) The OAM value increases by 1 with time or distance. (b) The OAM value decreases by 1 with time or distance. (c) The OAM value increases by 1 at $z = 5$ cm, decreases by 1 at $z = 15$ cm, and oscillates at $z = 25$ cm. (d) The OAM value changes discontinuously between neighboring times and distances.

**Discussion**

In our paper, we propose and demonstrate a new type of STWP with both temporally and longitudinally pre-engineered spatiotemporal evolution. Instead of previously published approaches, our full 2-D spectral manipulation offers wide flexibility in tailoring the temporal and longitudinal response of these STWPs. Consequently, different light DOFs can be independently controlled, including intensity, polarization, and transverse spatial distribution (e.g., OAM). Furthermore, we experimentally synthesize wave packets carrying TLV-OAM, thus providing a proof-of-concept demonstration of the underlying theory.

We note that there are various limitations in our current experimental setup, including: (a) *Synthesis bandwidth*: We spatially split a few temporal frequency lines and modulate the transverse field of each frequency at different locations of the SLM. Since different frequencies share the same SLM in the experiment, the size of the SLM limits the number of temporal frequencies to be manipulated, as well as the spatial resolution of the transverse field modulation [44]; and (b) *Polarization*: Our experimental scheme can only generate wave packets with a single polarization as the SLM is polarization dependent [44]. We believe that more advanced spatiotemporal synthesis schemes [21,22,31,45] can improve the performance of our approach.

Moreover, we chose to use the Bessel mode basis set in our approach for temporal and longitudinal evolution. These modes enable direct control of polarization and OAM of the transverse fields [41]. However, we believe our approach can potentially utilize other modal sets that have tunability in longitudinal wavenumbers [46–48]; different modal sets can have various transverse properties, and might be utilized to design temporally and longitudinally structured wave packets. Furthermore, we believe that our approach can be extended beyond scalar fields to vector fields, and may potentially enable temporal and longitudinal control of the polarization distribution of the transverse fields [31,37,49].

Additionally, we showed a dynamic transverse spatial distribution only in terms of the OAM value. Similar to combining multiple modes to design transverse intensity and phase profiles of monochromatic beams, we believe that our approach can also be extended to generate STWPs with devised transverse spatial distributions that evolve both temporally and longitudinally [4,21,31].

**Methods**

***Fourier synthesis of temporal and longitudinal waveform:*** For the 1-D case, temporal/longitudinal waveforms can be generated by the Fourier relation between the waveform and its corresponding temporal/longitudinal wavenumber spectrum. We use an equally spaced temporal and longitudinal comb to synthesize the desired temporally or longitudinally structured light field, and the selected wavenumbers can be given as

$$k_{\omega_n} = k_{\omega_0} + \frac{2\pi n}{T \cdot c}; \text{ or } k_{z_m} = k_{z_0} + \frac{2\pi m}{L} \tag{7}$$

where $k_{\omega_n}$ and $k_{z_m}$ are $n$-th temporal wavenumber and $m$-th longitudinal wavenumber, respectively; and $k_{\omega_0}$ and $k_{z_0}$ are temporal and longitudinal carrier-envelope offset wavenumbers, respectively. To obtain the devised waveform, different transformations can be utilized based on different scenarios (e.g., discrete/continuous spectrum, discrete/continuous waveform). Considering the discrete frequency spectrum and continuous waveform studied here, we perform the 1-D Fourier series to obtain the required complex coefficients at each frequency [34,42,43]:

$$A_n = \frac{1}{T}\int_0^T f(t) e^{-i(-\frac{2\pi}{T}n)t} dt; \text{ or } B_m = \frac{1}{L}\int_0^L g(z) e^{-i\frac{2\pi}{L}mz} dz \tag{8}$$

where $f(t)$ and $g(z)$ define the desired temporal and longitudinal complex waveform, respectively. In these two Fourier series equations, there are opposite signs on the exponential term due to the sign of phase terms for the temporal and longitudinal components in the harmonic waves. A perfect reconstruction of the arbitrary waveform theoretically requires infinite frequencies. We truncate the spectrum and obtain finite number of spectral components as an approximation.

To tailor the 2-D waveform in both the temporal and longitudinal domains, a 2-D spectral space consisting of both temporal and longitudinal wavenumbers is required. In our demonstration, the selection of the longitudinal spectrum depends on the temporal frequencies. We set a linear dispersion relation between the central longitudinal frequency and the temporal frequency; thus, the central value of $k_{z_{n,m}}$ at $n$-th temporal frequency is set as $\alpha \cdot k_{\omega_n}$, where $\alpha$ is a constant for all temporal frequencies. Correspondingly, the group velocity of the wave packet can be estimated as $v_g = \alpha \cdot c$. The 2-D spectral components are obtained as

$$k_{\omega_{n,m}} = k_{\omega_0} + \frac{2\pi n}{T \cdot c}; \ k_{z_{n,m}} = \alpha \cdot k_{\omega_{n,m}} + \frac{2\pi m}{L} \tag{9}$$

In our demonstration, $\alpha$ is set to be 0.999985, 0.99998, 0.99996, and 0.99987 when 6, 15, 31, and 101 temporal frequencies are employed, respectively. With a Galilean transformation, we design the waveform in the $z - \tau$ coordinates. The complex coefficients can be calculated from a 2-D Fourier series as follows:

$$C_{n,m} = \frac{1}{T}\frac{1}{L}\int_0^T \int_0^L s(\tau,z) e^{i\frac{2\pi}{T}n\tau} e^{-i\frac{2\pi}{L}mz} d\tau\, dz \tag{10}$$

where $s(\tau,z)$ is the function that describes the amplitude and phase waveform in terms of time and distance. Specifically, when the 2-D waveform is separable in the temporal and longitudinal domain as $s(\tau,z) = f(\tau)g(z)$, the Fourier series can be directly obtained as $C_{n,m} = A_n B_m$.

*Experimental details:* As shown in *Supplementary Note 2*, the output of the micro-ring resonator-based Kerr comb is fed into a programmable LCoS filter. Six temporal frequency lines are subsequently selected, and the amplitude and phase of the different frequencies are tuned to be the same by the programmable LCoS filter. The spectrum and autocorrelation results are presented in *Supplementary Note 2*. The optical pulse with an initial Gaussian spatial profile is divided into two paths for STWP generation and reference. To synthesize the designed light field, we first use two cascaded gratings to sperate the frequencies by a proper spatial spacing with a propagation distance of approximately 2 m. Lenses are used to form the beams in two rows, and 6 beams are projected on SLM 1 as a 3×2 array. We further generate digital holographic patterns on the SLM to directly modulate the intensity and phase profiles of each temporal frequency. Each transverse field is a coherent sum of 11 (longitudinal wavenumbers) × 4 (time slots) × 3 (longitudinal slots) components. The reflecting angle of each beam is tuned, such that all beams arrive at the same position as that of SLM 2. Six grating patterns are combined on SLM 2. Each grating tailors the reflection angle of each temporal frequency, and all 6 frequency components propagate coaxially after SLM 2. Our frequency combining method will induce a $1/N$ loss when combining $N$ beams. A 2-lens 4-$f$ imager system is used to relay the field generated at SLM 1 to the observation plane. The focal length of the lens is set as 80 cm. The STWP is then combined with the Gaussian reference pulse at the center of the camera while arriving from a different angle. Two translation delay stages are utilized in both the STWP synthesizing path and the reference path. The initial position of each stage is carefully aligned. We place the camera at the $z = 0$ plane and temporally align the reference pulse with the STWP when both delays are set to 0.

*Spatiotemporal off-axis hologram:* To reconstruct the total electric field of the STWP at different time instants and axial distances, we perform an off-axis holography and sweep the delay stage on the wave packet arm ($\Delta z_{STWP}$) and the reference arm ($\Delta z_{ref}$) [50]. The time-averaged intensity profile recorded by the camera can be depicted as follows:

$$I_{camera}(x,y;\Delta z_{STWP},\Delta z_{ref}) \propto \left|\sum_n E_{STWP,n}(x,y;\Delta z_{STWP}) + E_{ref,n}(x,y;\Delta z_{ref}) \cdot e^{ik_x x}\right|^2 \tag{11}$$

where $E_{STWP,n}(x,y;\Delta z_{STWP})$ and $E_{ref,n}(x,y;\Delta z_{ref})$ are the electric fields of the wave packet to be measured and the reference pulse at the $n$-th temporal frequency at the distance set by the delay stage

$\Delta z_{STWP}$ and $\Delta z_{ref}$, respectively; and $e^{ik_x x}$ represents the spatial phase difference produced by different arriving angles of the STWP and the reference. Since the reference pulse has a relatively large Gaussian spatial profile, we assume it to be a series of plane waves with different frequencies propagating approximately at the speed of $c$. Equation (12) shows the estimated electric field of the reference pulse at the $n$-th temporal frequency:

$$E_{ref,n}(x, y; \Delta z_{ref}) \propto e^{ik_{zref,n}\Delta z_{ref}} = e^{-ik_{\omega_n} c(-\Delta z_{ref}/c)} \quad (12)$$

Since the camera has a limited bandwidth, cross-temporal frequency beatings are not recorded in the picture. Thus, the recorded intensity profile can be simplified as follows:

$$I_{camera} \propto \sum_n |E_{STWP,n}|^2 + \sum_n |E_{ref,n}|^2 + \sum_n E_{STWP,n}(E_{ref,n}e^{ik_x x})^* + \sum_n E_{STWP,n}^* E_{ref,n} e^{ik_x x} \quad (13)$$

Performing a Fourier transform and spatial frequency filtering-based image processing method (see *Supplementary Note 3* for more details), we can retrieve the electric field from the off-axis holograph as:

$$E_{off-axis}(x, y; \Delta z_{STWP}, \Delta z_{ref}) = \sum_n E_{STWP,n}(x, y; \Delta z_{STWP}) E_{ref,n}(x, y; \Delta z_{ref})^*$$
$$= \sum_n E_{STWP,n}(x, y; \Delta z_{STWP}) e^{-ik_{\omega_n} c(\Delta z_{ref}/c)} \quad (14)$$

The translation delay produced by the reference can be regarded as time variable $t = \Delta z_{ref}/c$ for the wave packet. Therefore, when using the time axis $\tau = t - z/v_g$, the right side of Equation (15) can be used to estimate the electric field of the STWP at a specific time and distance:

$$E_{STWP}(x, y; z, \tau) = E_{off-axis}(x, y; z = \Delta z_{STWP}, \tau = \Delta z_{ref}/c - \Delta z_{STWP}/v_g) \quad (15)$$

## Reference

[1] Auston D 1968 Transverse mode locking *IEEE Journal of Quantum Electronics* **4** 420–2
[2] Yessenov M, Hall L A, Schepler K L and Abouraddy A F 2022 Space-time wave packets *Adv. Opt. Photon.* **14** 455–570
[3] Jolly S W, Gobert O and Quéré F 2020 Spatio-temporal characterization of ultrashort laser beams: a tutorial *J. Opt.* **22** 103501
[4] Pierce J R, Palastro J P, Li F, Malaca B, Ramsey D, Vieira J, Weichman K and Mori W B 2023 Arbitrarily structured laser pulses *Phys. Rev. Res.* **5** 013085
[5] Shen Y, Zhan Q, Wright L G, Christodoulides D N, Wise F, Willner A, Zou K, Zhao Z, Porras M A, Chong A, Wan C, Bliokh K Y, Liao C-T, Hernandez Garcia C, Murnane M M, Yessenov M, Abouraddy A, Wong L J, GO M, Kumar S, Guo C, Fan S, Papasimakis N, Zheludev N I, Chen L, Zhu W, Agrawal A, Mounaix M, Fontaine N K, Carpenter J, Jolly S W, Dorrer C, Alonso B, Lopez-Quintas I, López-Ripa M, Sola I, Huang J, Zhang H, Ruan Z, Dorrah A H, Capasso F and Forbes A 2023 Roadmap on spatiotemporal light fields *J. Opt.*
[6] Krupa K, Tonello A, Shalaby B M, Fabert M, Barthélémy A, Millot G, Wabnitz S and Couderc V 2017 Spatial beam self-cleaning in multimode fibres *Nat. Photon.* **11** 237–41
[7] Wright L G, Wu F O, Christodoulides D N and Wise F W 2022 Physics of highly multimode nonlinear

optical systems *Nat. Phys.* **18** 1018–30
[8]  Divitt S, Zhu W, Zhang C, Lezec H J and Agrawal A 2019 Ultrafast optical pulse shaping using dielectric metasurfaces *Science* **364** 890–4
[9]  Chong A, Renninger W H, Christodoulides D N and Wise F W 2010 Airy–Bessel wave packets as versatile linear light bullets *Nat. Photon.* **4** 103–6
[10]  Bhaduri B, Yessenov M and Abouraddy A F 2020 Anomalous refraction of optical spacetime wave packets *Nat. Photon.* **14** 416–21
[11]  Yessenov M, Free J, Chen Z, Johnson E G, Lavery M P J, Alonso M A and Abouraddy A F 2022 Space-time wave packets localized in all dimensions *Nat. Commun.* **13** 4573
[12]  Yessenov M and Abouraddy A F 2020 Accelerating and decelerating space-time optical wave packets in free space *Phys. Rev. Lett.* **125** 233901
[13]  Froula D H, Turnbull D, Davies A S, Kessler T J, Haberberger D, Palastro J P, Bahk S-W, Begishev I A, Boni R, Bucht S, Katz J and Shaw J L 2018 Spatiotemporal control of laser intensity *Nat. Photon.* **12** 262–5
[14]  Turnbull D, Bucht S, Davies A, Haberberger D, Kessler T, Shaw J L and Froula D H 2018 Raman Amplification with a Flying Focus *Phys. Rev. Lett.* **120** 024801
[15]  Zhao Z, Song H, Zhang R, Pang K, Liu C, Song H, Almaiman A, Manukyan K, Zhou H, Lynn B, Boyd R W, Tur M and Willner A E 2020 Dynamic spatiotemporal beams that combine two independent and controllable orbital-angular-momenta using multiple optical-frequency-comb lines *Nat. Commun.* **11** 4099
[16]  Béjot P and Kibler B 2021 Spatiotemporal helicon wavepackets *ACS Photonics* **8** 2345–54
[17]  Piccardo M, de Oliveira M, Policht V R, Russo M, Ardini B, Corti M, Valentini G, Vieira J, Manzoni C, Cerullo G and Ambrosio A 2023 Broadband control of topological–spectral correlations in space–time beams *Nat. Photon.* 1–7
[18]  Vieira J, Mendonça J T and Quéré F 2018 Optical Control of the Topology of Laser-Plasma Accelerators *Phys. Rev. Lett.* **121** 054801
[19]  Rego L, Dorney K M, Brooks N J, Nguyen Q L, Liao C-T, San Román J, Couch D E, Liu A, Pisanty E, Lewenstein M, Plaja L, Kapteyn H C, Murnane M M and Hernández-García C 2019 Generation of extreme-ultraviolet beams with time-varying orbital angular momentum *Science* **364** eaaw9486
[20]  Zou K, Su X, Yessenov M, Pang K, Karapetyan N, Karpov M, Song H, Zhang R, Zhou H, Kippenberg T J, Tur M, Abouraddy A F, Willner A E and Willner A E 2022 Tunability of space-time wave packet carrying tunable and dynamically changing OAM value *Opt. Lett.* **47** 5751–4
[21]  Cruz-Delgado D, Yerolatsitis S, Fontaine N K, Christodoulides D N, Amezcua-Correa R and Bandres M A 2022 Synthesis of ultrafast wavepackets with tailored spatiotemporal properties *Nat. Photon.* **16** 686–91
[22]  Chen L, Zhu W, Huo P, Song J, Lezec H J, Xu T and Agrawal A 2022 Synthesizing ultrafast optical pulses with arbitrary spatiotemporal control *Science Advances* **8** eabq8314
[23]  Chong A, Wan C, Chen J and Zhan Q 2020 Generation of spatiotemporal optical vortices with controllable transverse orbital angular momentum *Nat. Photon.* **14** 350–4
[24]  Jhajj N, Larkin I, Rosenthal E W, Zahedpour S, Wahlstrand J K and Milchberg H M 2016 Spatiotemporal optical vortices *Phys. Rev. X* **6** 031037
[25]  Hancock S W, Zahedpour S and Milchberg H M 2021 Second-harmonic generation of spatiotemporal optical vortices and conservation of orbital angular momentum *Optica* **8** 594–7
[26]  Gui G, Brooks N J, Kapteyn H C, Murnane M M and Liao C-T 2021 Second-harmonic generation and the conservation of spatiotemporal orbital angular momentum of light *Nat. Photon.* **15** 608–13
[27]  Zdagkas A, McDonnell C, Deng J, Shen Y, Li G, Ellenbogen T, Papasimakis N and Zheludev N I 2022 Observation of toroidal pulses of light *Nat. Photon.* **16** 523–8
[28]  Wan C, Cao Q, Chen J, Chong A and Zhan Q 2022 Toroidal vortices of light *Nat. Photon.* **16** 519–22
[29]  Yessenov M, Hall L A, Ponomarenko S A and Abouraddy A F 2020 Veiled talbot effect *Phys. Rev. Lett.* **125** 243901
[30]  Guo C, Xiao M, Orenstein M and Fan S 2021 Structured 3D linear space–time light bullets by nonlocal

    nanophotonics *Light Sci. Appl.* **10** 160
[31] Mounaix M, Fontaine N K, Neilson D T, Ryf R, Chen H, Alvarado-Zacarias J C and Carpenter J 2020 Time reversed optical waves by arbitrary vector spatiotemporal field generation *Nat. Commun.* **11** 5813
[32] Forbes A, de Oliveira M and Dennis M R 2021 Structured light *Nat. Photon.* **15** 253–62
[33] Vieira T A, Gesualdi M R R and Zamboni-Rached M 2012 Frozen waves: experimental generation *Opt. Lett.* **37** 2034–6
[34] Zamboni-Rached M and Mojahedi M 2015 Shaping finite-energy diffraction- and attenuation-resistant beams through Bessel-Gauss--beam superposition *Phys. Rev. A* **92** 043839
[35] Dorrah A H, Rubin N A, Zaidi A, Tamagnone M and Capasso F 2021 Metasurface optics for on-demand polarization transformations along the optical path *Nat. Photon.* **15** 287–96
[36] Moreno I, Davis J A, Sánchez-López M M, Badham K and Cottrell D M 2015 Nondiffracting Bessel beams with polarization state that varies with propagation distance *Opt. Lett.* **40** 5451–4
[37] Otte E, Rosales-Guzmán C, Ndagano B, Denz C and Forbes A 2018 Entanglement beating in free space through spin–orbit coupling *Light Sci. Appl.* **7** 18009–18009
[38] Dorrah A H, Zamboni-Rached M and Mojahedi M 2016 Controlling the topological charge of twisted light beams with propagation *Phys. Rev. A* **93** 063864
[39] Dorrah A H, Rubin N A, Tamagnone M, Zaidi A and Capasso F 2021 Structuring total angular momentum of light along the propagation direction with polarization-controlled meta-optics *Nat. Commun.* **12** 6249
[40] Saleh B E A and Teich M C 2019 *Fundamentals of Photonics* (John Wiley & Sons)
[41] McGloin D and Dholakia K 2005 Bessel beams: Diffraction in a new light *Contemporary Physics* **46** 15–28
[42] Weiner A M 2011 Ultrafast optical pulse shaping: A tutorial review *Opt. Commun.* **284** 3669–92
[43] Cundiff S T and Weiner A M 2010 Optical arbitrary waveform generation *Nat. Photon.* **4** 760–6
[44] Rosales-Guzmán C and Forbes A 2017 *How to shape light with spatial light modulators* (SPIE PRESS)
[45] Fontaine N K, Ryf R, Chen H, Neilson D T, Kim K and Carpenter J 2019 Laguerre-Gaussian mode sorter *Nat. Commun.* **10** 1865
[46] Zannotti A, Denz C, Alonso M A and Dennis M R 2020 Shaping caustics into propagation-invariant light *Nat. Commun.* **11** 3597
[47] Lóxpez-Mariscal C, Gutiérrez-Vega J C, Milne G and Dholakia K 2006 Orbital angular momentum transfer in helical Mathieu beams *Opt. Express* **14** 4182–7
[48] Dudley A, Vasilyeu R, Belyi V, Khilo N, Ropot P and Forbes A 2012 Controlling the evolution of nondiffracting speckle by complex amplitude modulation on a phase-only spatial light modulator *Opt. Commun.* **285** 5–12
[49] Yessenov M, Chen Z, Lavery M P J and Abouraddy A F 2022 Vector space-time wave packets *Opt. Lett.* **47** 4131–4
[50] Cuche E, Marquet P and Depeursinge C 2000 Spatial filtering for zero-order and twin-image elimination in digital off-axis holography *Appl. Opt.* **39** 4070–5

**Acknowledgments**

We thank Prof. Ayman F. Abouraddy and Dr. Murat Yessenov for discussions. This work is supported by the Office of Naval Research through a MURI Award N00014-20-1-2789; Defense University Research Instrumentation Program (DURIP) (FA9550-20-1-0152); and Qualcomm Innovation Fellowship.

**Author contributions**

X.S. conceived the idea; X.S. and K.Z. designed and carried out the experiment with the help of H.Z., H.S., and Y.D.; X.S., K.Z., H.Z., H.S., and Z.J. developed the theory and simulation model; X.S., Y.W., R.Z., and Y.D. carried out data analysis; M.K. and T.J.K. provided the Kerr comb chip; T.J.K., M.T., D.N.C., and A.E.W. provided the technical support. All the authors contributed to the interpretation of the results and manuscript writing. The project was supervised by A.E.W.

# Extended Data Figures 1-5

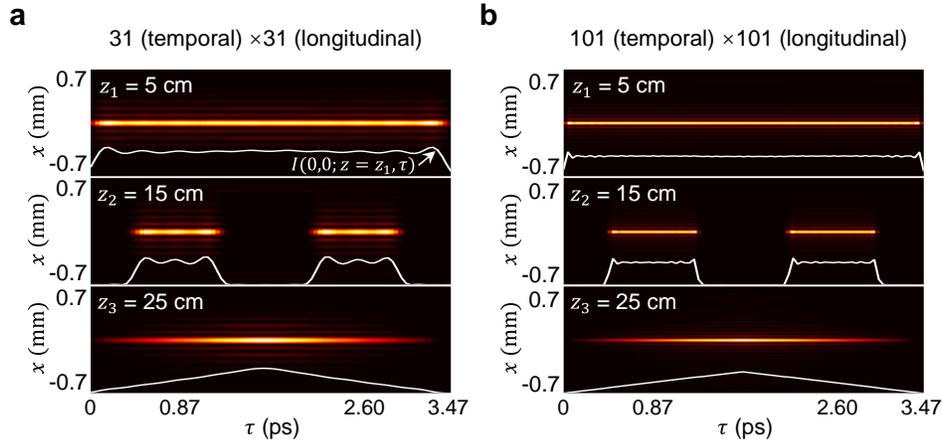

**Extended Data Fig. 1. Simulated the intensity profiles along $x$-axis for STWP with temporally and longitudinally structured waveform.** These two wave packets correspond to the cases shown in Fig.3(c). Different numbers of temporal and longitudinal spectral components are utilized to construct the same ideal waveform design: (a) 31 temporal and 31 longitudinal wavenumbers; and (b) 101 temporal and 101 longitudinal wavenumbers. The solid white line shows the on-axis intensity. With an increase in the number of frequencies, the generated wave packets have a closer on-axis waveform compared to the designed waveform.

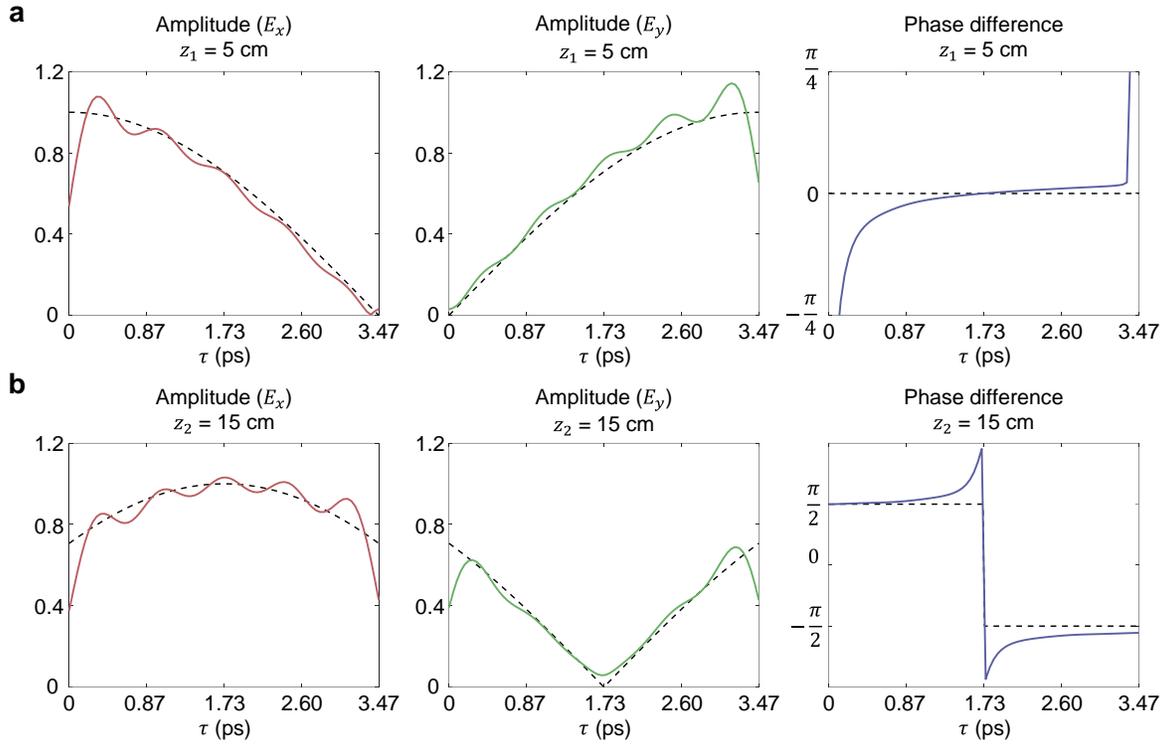

**Extended Data Fig. 2. Simulated on-axis waveform of an STWP with time- and longitudinal-varying polarization.** Presented waveforms correspond to the same wave packet in Fig. 3(d). Ideal and simulated amplitude waveforms of $E_x$ and $E_y$, and phase difference $\varphi(E_y) - \varphi(E_x)$ are shown at (a) $z = 5$ cm and (b) $z = 15$ cm.

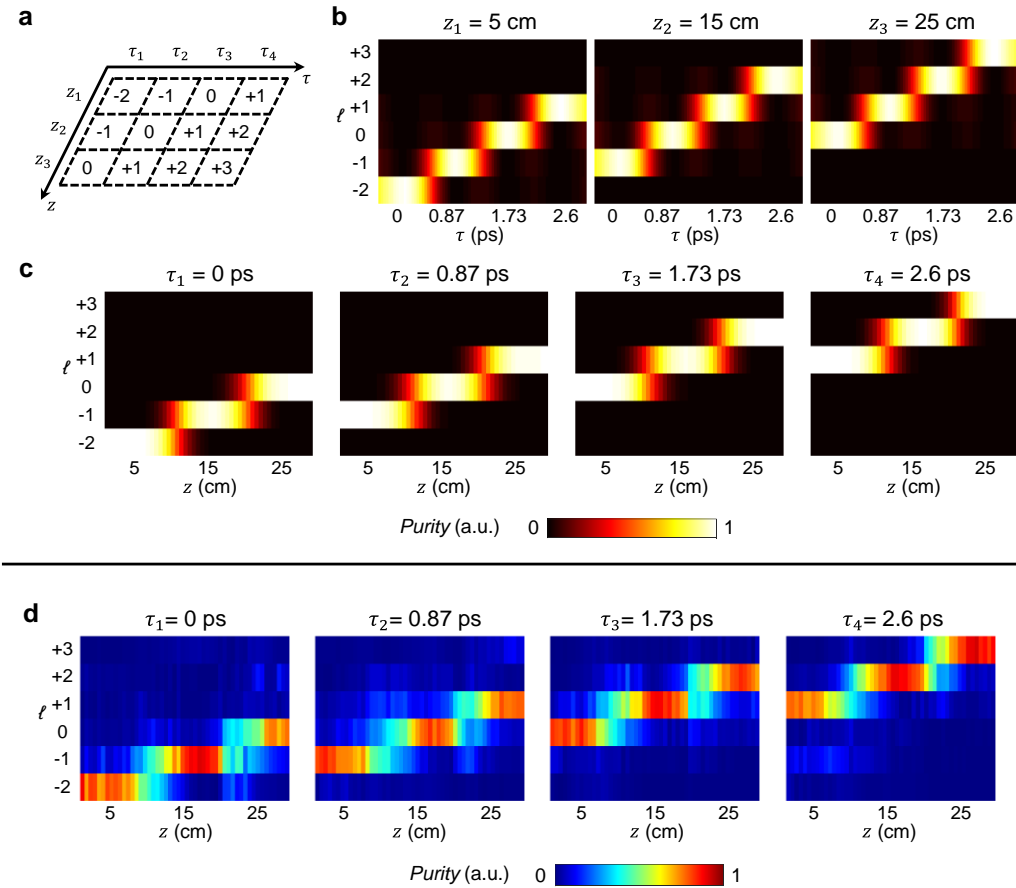

**Extended Data Fig. 3. Color coded modal spectra for TLV-OAM wave packets.** (a) Mode assignment for the wave packet. The OAM value increases by 1 with in time or distance. (b) Simulated OAM modal spectra as a function of time at three central distances. (c) Simulated OAM modal spectra as a function of distance at four central time instants. (d) Experimental OAM modal spectra as a function of distance at four central time instants.

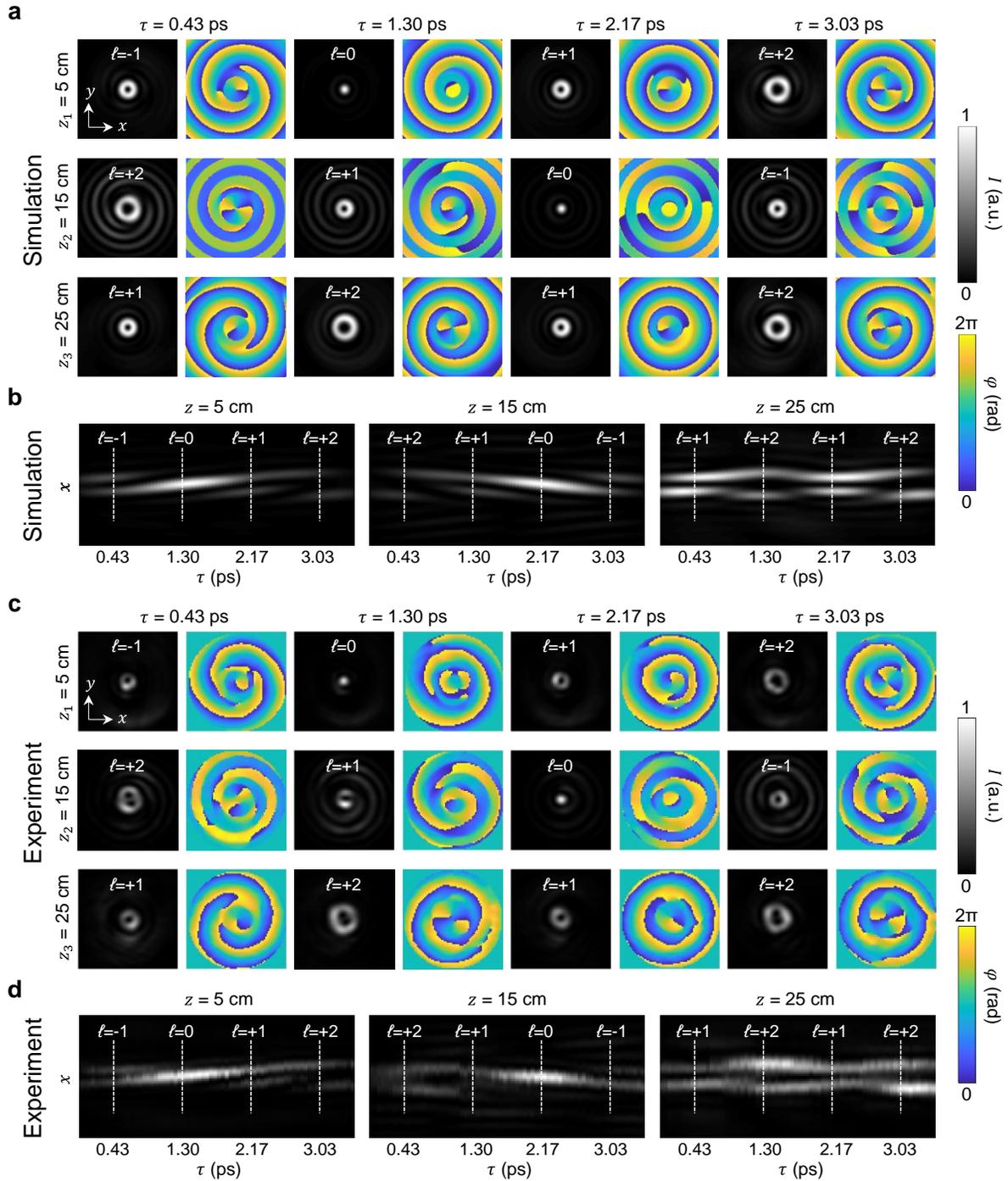

**Extended Data Fig. 4. Profiles of a STWP carrying TLV-OAM.** Transverse profiles for the wave packet shown in Fig. 5(c). The OAM value increases by 1 at $z=5$ cm, decreases by 1 at $z=15$ cm, and oscillates at $z=25$ cm. (a) Simulated intensity and phase profiles at different time instants and distances. (b) Temporal evolution of intensity profiles along $x$-axis $I(x, y=0; z=z_i, \tau)$ at three distances. (c) Experimentally measured intensity and phase profiles of the same TLV-OAM wave packet. (d) Temporal evolution of experimental intensity profiles along $x$-axis.

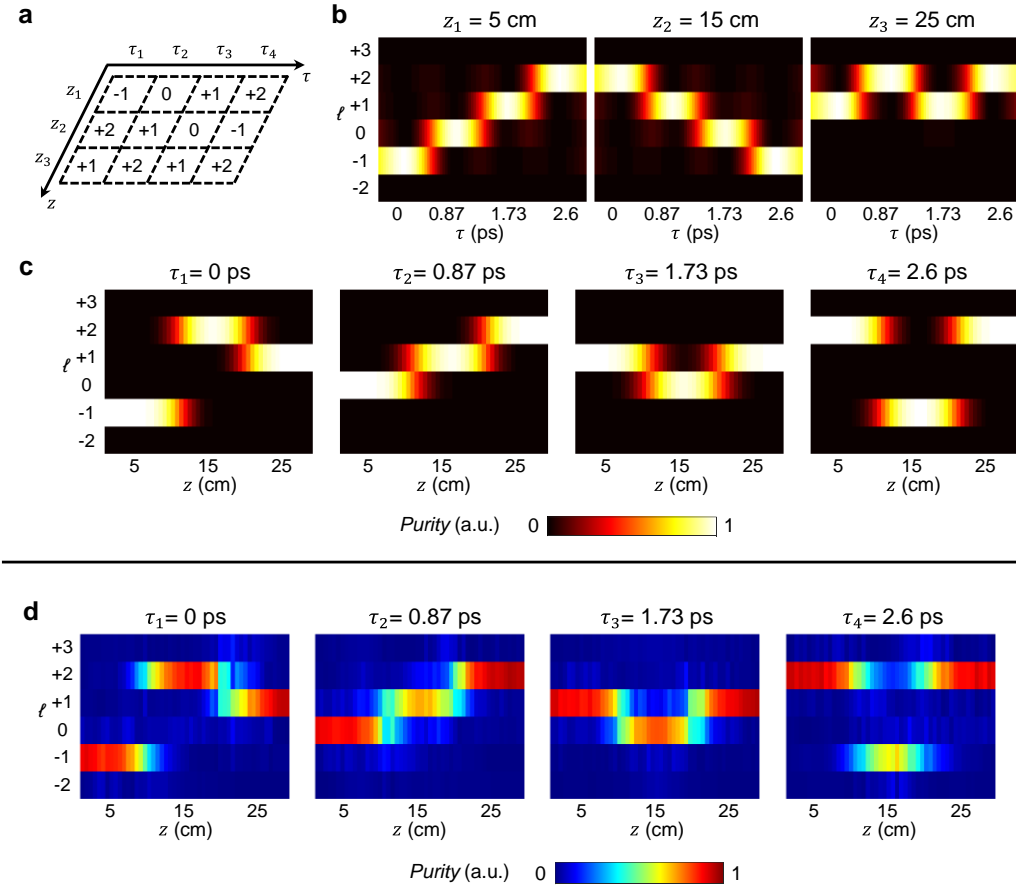

**Extended Data Fig. 5. Color coded modal spectra for TLV-OAM wave packets.** (a) Mode assignment for the wave packet. The OAM value increases by 1 at $z=5$ cm, decreases by 1 at $z=15$ cm, and oscillates at $z=25$ cm. (b) Simulated OAM modal spectra as a function of time at three central distances. (c) Simulated OAM modal spectra as a function of distance at four central time instants. (d) Experimental OAM modal spectra as a function of distance at four central time instants.